\documentclass[12pt,twoside]{article}
\usepackage{latexsym}
\def\eqvsp{}  \newdimen\paravsp  \paravsp=1.3ex
\def\,{\mskip 3mu} \def\>{\mskip 4mu plus 2mu minus 4mu} \def\;{\mskip 5mu plus 5mu} \def\!{\mskip-3mu}
\def\dispmuskip{\thinmuskip= 3mu plus 0mu minus 2mu \medmuskip=  4mu plus 2mu minus 2mu \thickmuskip=5mu plus 5mu minus 2mu}
\def\textmuskip{\thinmuskip= 0mu                    \medmuskip=  1mu plus 1mu minus 1mu \thickmuskip=2mu plus 3mu minus 1mu}
\textmuskip
\def\beq{\eqvsp\dispmuskip\begin{equation}}
\def\eeq{\eqvsp\end{equation}\textmuskip}
\def\beqn{\eqvsp\dispmuskip\begin{displaymath}}\def\eeqn{\eqvsp\end{displaymath}\textmuskip}
\def\bqa{\eqvsp\dispmuskip\begin{eqnarray}}
\def\eqa{\eqvsp\end{eqnarray}\textmuskip}
\def\bqan{\eqvsp\dispmuskip\begin{eqnarray*}}
\def\eqan{\eqvsp\end{eqnarray*}\textmuskip}
\newenvironment{keywords}{\centerline{\bf\small Keywords}\begin{quote}\small}{\par\end{quote}\vskip 1ex}
\def\paradot#1{\vspace{\paravsp plus 0.5\paravsp minus 0.5\paravsp}\noindent{\bf\boldmath{#1.}}}
\def\paranodot#1{\vspace{\paravsp plus 0.5\paravsp minus 0.5\paravsp}\noindent{\bf\boldmath{#1}}}
\def\req#1{(\ref{#1})}
\def\epstr{\epsilon}
\def\fr#1#2{{\textstyle{#1\over#2}}}
\def\SetR{I\!\!R}
\def\SetN{I\!\!N}
\def\qmbox#1{{\quad\mbox{#1}\quad}}
\def\e{{\rm e}}
\def\lb{{\log_2}}
\def\UTM{\mbox{UTM}}
\def\Length{\mbox{Length}}
\def\gtrsim{\buildrel{\lower.7ex\hbox{$>$}}\over{\lower.7ex\hbox{$\sim$}}}

\begin{document}
\def\baselinestretch{0.95}

\title{\vspace{-4ex}
\vskip 2mm\bf\Large\hrule height5pt \vskip 2mm
A Complete Theory of Everything \\ (will be subjective)
\vskip 2mm \hrule height2pt}
\author{{\bf Marcus Hutter}\\[1mm]
\normalsize SoCS, RSISE, IAS, CECS \\[-0.5ex] 
\normalsize Australian National University \\[-0.5ex]
\normalsize Canberra, ACT, 0200, Australia \\
}
\date{September 2010}
\maketitle
\vspace*{-7ex}

\begin{abstract}
Increasingly encompassing models have been suggested for our world.
Theories range from generally accepted to increasingly
speculative to apparently bogus. The progression of theories from
ego- to geo- to helio-centric models to universe and multiverse
theories and beyond was accompanied by a dramatic increase in the
sizes of the postulated worlds, with humans being expelled from
their center to ever more remote and random locations.
Rather than leading to a true theory of everything, this trend faces
a turning point after which the predictive power of such theories
decreases (actually to zero).
Incorporating the location and other capacities of the observer into
such theories avoids this problem and allows to distinguish
meaningful from predictively meaningless theories.
This also leads to a truly complete theory of everything consisting
of a (conventional objective) theory of everything plus a (novel
subjective) observer process.
The observer localization is neither based on the controversial
anthropic principle, nor has it anything to do with the
quantum-mechanical observation process.
The suggested principle is extended to more practical (partial,
approximate, probabilistic, parametric) world models (rather than
theories of everything).
Finally, I provide a justification of Ockham's razor, and criticize
the anthropic principle, the doomsday argument, the no free lunch
theorem, and the falsifiability dogma.
\def\contentsname{\centering\normalsize Contents\vspace{0ex}}
{\parskip=-2.7ex\tableofcontents}
\end{abstract}

\vspace*{-4ex}

\begin{quote}\it
``... in spite of it's incomputability, Algorithmic Probability can
serve as a kind of `Gold Standard' for induction systems'' \par
\hfill --- {\sl Ray Solomonoff (1997)}
\end{quote}

\begin{quote}\it
``There is a theory which states that if ever anyone discovers exactly what
the Universe is for and why it is here, it will instantly disappear and be
replaced by something even more bizarre and inexplicable. There is another
theory which states that this has already happened.'' \par
\hfill --- {\sl Douglas Adams, Hitchhikers guide to the Galaxy (1979)}
\end{quote}

\section{Introduction}\label{secIntro}

This paper uses an {\em information-theoretic} and {\em
computational} approach for addressing the {\em philosophical}
problem of judging theories (of everything) in {\em physics}. In
order to keep the paper generally accessible,
I've tried to minimize field-specific jargon and mathematics,
and focus on the core problem and its solution.

By {\em theory} I mean any {\em model} which can
explain$\approx$describe$\approx$predict$\approx$compress
\cite{Hutter:06hprize} our observations, whatever the form of the
model. Scientists often say that their model {\em explains} some
phenomenon. What is usually meant is that the model {\em describes}
(the relevant aspects of) the observations more compactly than the
raw data. The model is then regarded as capturing a law (of nature),
which is believed to hold true also for unseen/future data.

This process of inferring general conclusions from example instances
is call {\em inductive reasoning}. For instance, observing 1000
black ravens but no white one supports but cannot prove the
hypothesis that all ravens are black.
In general, induction is used to find properties or rules or models
of past observations. The ultimate purpose of the induced models is
to use them for making predictions, e.g.\ that the next observed
raven will also be black.
Arguably inductive reasoning is even more important than deductive
reasoning in science and everyday life:
for scientific discovery, in machine learning, for forecasting in
economics, as a philosophical discipline, in common-sense decision
making, and last but not least to find theories of everything.
Historically, some famous, but apparently misguided philosophers
\cite{Stove:82,Gardner:01}, including Popper and Miller, even
disputed the existence, necessity or validity of inductive
reasoning.
Meanwhile it is well-known how minimum encoding length principles
\cite{Wallace:05,Gruenwald:07book}, rooted in (algorithmic)
information theory \cite{Li:08}, quantify Ockham's razor principle,
and lead to a solid pragmatic foundation of inductive reasoning
\cite{Hutter:07uspx}.
Essentially, one can show that the more one can {\em compress}, the
better one can {\em predict}, and vice versa.

A deterministic theory/model allows from initial conditions to
determine an observation sequence, which could be coded as a bit
string. For instance, Newton mechanics maps initial planet
positions+velocities into a time-series of planet positions. So a
deterministic model with initial conditions is ``just'' a compact
representation of an infinite observation string. A stochastic model
is ``just'' a probability distribution over observation strings.

Classical models in physics are essentially differential equations
describing the time-evolution of some aspects of the world. A Theory
of Everything (ToE) models the whole universe or multiverse, which
should include initial conditions. As I will argue, it can be
crucial to also localize the observer, i.e.\ to augment the ToE with
a model of the properties of the observer, even for
non-quantum-mechanical phenomena. I call a ToE with observer
localization, a {\em Complete ToE} (CToE).

That the observer itself is important in describing our world is
well-known. Most prominently in quantum mechanics, the observer
plays an active role in `collapsing the wave function'$\!$. This is
a specific and relatively well-defined role of the observer for a
particular theory, which is {\em not} my concern. I will show that
(even the localization of) the observer is indispensable for {\em
finding} or developing {\em any} (useful) ToE.
Often, the anthropic principle is invoked for this purpose (our
universe is as it is because otherwise we would not exist).
Unfortunately its current use is rather vague and limited, if not
outright unscientific \cite{Smolin:04}.
In Section \ref{secUToE} I extend Schmidhuber's formal work
\cite{Schmidhuber:00toe} on computable ToEs to {\em formally}
include observers. Schmidhuber \cite{Schmidhuber:00toe} already
discusses observers and mentions sampling universes consistent with
our own existence, but this part stays informal.
I give a precise and formal account of observers by explicitly
separating the observer's subjective experience from the objectively
existing universe or multiverse, which besides other things shows
that we also need to localize the observer within our universe (not
only which universe the observer is in).

In order to make the main point of this paper clear, Section
\ref{secSToE} first traverses a number of models that have been
suggested for our world, from generally accepted to increasingly
speculative and questionable theories.
Section \ref{secPPOL} discusses the relative merits of the models, in
particular their predictive power (precision and coverage). We will
see that localizing the observer, which is usually not regarded as
an issue, can be very important.
Section \ref{secCToEs} gives an informal introduction to the
necessary ingredients for CToEs, and how to evaluate and compare
them using a quantified instantiation of Ockham's razor.
Section \ref{secCToE} gives a formal definition of what accounts for
a CToE, introduces more realistic observers with limited perception
ability, and formalizes the CToE selection principle.
The Universal ToE is a sanity critical point in the development of
ToEs, and will be investigated in more detail in Section \ref{secUToE}.
Extensions to more practical (partial, approximate, probabilistic, parametric)
theories (rather than ToEs) are briefly discussed in Section \ref{secExt}.
In Section \ref{secPOR} I show that Ockham's razor is well-suited
for finding ToEs and briefly criticize the anthropic principle, the
doomsday argument, the no free lunch theorem, and the falsifiability
dogma.
Section \ref{secDisc} concludes.

\section{Theories of Something, Everything \& Nothing}\label{secSToE}

A number of models have been suggested for our world. They range
from generally accepted to increasingly speculative to outright
unacceptable. For the purpose of this work it doesn't matter where you
personally draw the line.
Many now generally accepted theories have once been regarded as
insane, so using the scientific community or general public as a
judge is problematic and can lead to endless discussions: for
instance, the historic geo$\leftrightarrow$heliocentric battle; and
the ongoing discussion of whether string theory is a theory of
everything or more a theory of nothing.
In a sense this paper is about a formal rational criterion to
determine whether a model makes sense or not.
In order to make the main point of this paper clear, below I will
briefly traverse a number of models. Space constraints prevent to
explain these models properly, but most of them are commonly known;
see e.g.\ \cite{Harrison:00,Barrow:04} for surveys. The presented
bogus models help to make clear the necessity of observer
localization and hence the importance of this work.

\paradot{(G) Geocentric model}
In the well-known geocentric model, the Earth is at the center of
the universe and the Sun, the Moon, and all planets and stars move
around Earth. The ancient model assumed concentric spheres, but
increasing precision in observations and measurements revealed a
quite complex geocentric picture with planets moving with variable
speed on epicycles. This Ptolemaic system predicted the celestial
motions quite well for its time, but was relatively complex in the
common sense and in the sense of involving many parameters that
had to be fitted experimentally.

\paradot{(H) Heliocentric model}
In the modern (later) heliocentric model, the Sun is at the
center of the solar system (or universe), with all planets (and
stars) moving in ellipses around the Sun. Copernicus developed a
complete model, much simpler than the Ptolemaic system, which
interestingly did not offer better predictions initially, but
Kepler's refinements ultimately outperformed all geocentric
models.
The price for this improvement was to expel the observers (humans)
from the center of the universe to one out of 8 moving planets.
While today this price seems small, historically it was quite high.
Indeed we will compute the exact price later.

\paradot{(E) Effective theories}
After the celestial mechanics of planets have been understood,
ever more complex phenomena could be captured with increasing
coverage.
Newton's mechanics unifies celestial and terrestrial gravitational
phenomena. When unified with special relativity theory one arrives
at Einstein's general relativity, predicting large scale phenomena
like black holes and the big bang.
On the small scale, electrical and magnetic phenomena are unified
by Maxwell's equations for electromagnetism.
Quantum mechanics and electromagnetism have further been unified to
quantum electrodynamics (QED). QED is the most powerful theory ever
invented, in terms of precision and coverage of phenomena. It is a
theory of all physical and chemical processes, except for
radio-activity and gravity.

\paradot{(P) Standard model of particle physics}
Salam, Glashow and Weinberg extended QED to include weak
interactions, responsible for radioactive decay. Together with
quantum chromo dynamic \cite{Hutter:96thesis}, which describes the
nucleus, this constitutes the current standard model (SM) of particle
physics. It describes all known non-gravitational phenomena in our
universe. There is no experiment indicating any limitation
(precision, coverage). It has about 20 unexplained parameters
(mostly masses and coupling constants) that have to be (and are)
experimentally determined (although some regularities can be explained
\cite{Hutter:97family}).
The effective theories of the previous paragraph can be regarded as
approximations of SM, hence SM, although founded on a subatomic
level, also predicts medium scale phenomena.

\paradot{(S) String theory}
Pure gravitational and pure quantum phenomena are perfectly
predictable by general relativity and the standard model,
respectively. Phenomena involving both, like the big bang, require a
proper final unification.
String theory is {\em the} candidate for a final unification of the
standard model with the gravitational force. As such it describes
the universe at its largest and smallest scale, and all scales
in-between. String theory is essentially parameter-free, but is
immensely difficult to evaluate and it seems to allow for many
solutions (spatial compactifications). For these and other reasons,
there is currently no uniquely accepted cosmological model.

\paradot{(C) Cosmological models}
Our concept of what the universe is, seems to ever expand. In
ancient times there was Earth, Sun, Moon, and a few planets,
surrounded by a sphere of shiny points (fixed stars). The current
textbook universe started in a big bang and
consists of billions of galaxy clusters
each containing billions of stars,
probably many with a planetary system.
But this is just the visible universe.
According to inflation models, which are needed to explain
the homogeneity of our universe, the ``total''
universe is vastly larger than the visible part.

\paradot{(M) Multiverse theories}
Many theories (can be argued to) imply a multitude of essentially
disconnected universes (in the conventional sense), often each
with their own (quite different) characteristics \cite{Tegmark:04}.
In Wheeler's oscillating universe a new big bang follows
the assumed big crunch, and this repeats indefinitely.
Lee Smolin proposed that every black hole recursively produces new
universes on the ``other side'' with quite different properties.
Everett's many-worlds interpretation of quantum mechanics postulates
that the wave function doesn't collapse but the universe splits
(decoheres) into different branches, one for each possible outcome
of a measurement.
Some string theorists have suggested that possibly all
compactifications in their theory are realized, each resulting in a
different universe.

\paradot{(U) Universal ToE}
The last two multiverse suggestions contain the seed of a general
idea. If theory X contains some unexplained elements Y (quantum or
compactification or other indeterminism), one postulates that every
realization of Y results in its own universe, and we just happen to
live in one of them. Often the anthropic principle is used in some
hand-waving way to argue why we are in this and not that universe
\cite{Smolin:04}.
Taking this to the extreme, Schmidhuber
\cite{Schmidhuber:97brauer,Schmidhuber:00toe} postulates a
multiverse (which I call universal universe) that consists of {\em
every} computable universe (note there are ``just'' countably many
computer programs). Clearly, if our universe is computable, then it
is contained in the universal universe, so we have a ToE already in
our hands. Similar in spirit but neither constructive nor formally
well-defined is Tegmark's mathematical multiverse \cite{Tegmark:08}.

\paradot{(R) Random universe}
Actually there is a much simpler way of obtaining a ToE. Consider an
infinite sequence of random bits (fair coin tosses). It is easy to
see that any finite pattern, i.e.\ any finite binary sequence,
occurs (actually infinitely often) in this string. Now consider our
observable universe quantized at e.g.\ Planck level, and code the
whole space-time universe into a huge bit string. If the universe
ends in a big crunch, this string is finite. (Think of a digital
high resolution 3D movie of the universe from the big bang to the
big crunch). This big string also appears somewhere in our random
string, hence our random string is a perfect ToE.
This is reminiscent of the Boltzmann brain idea that
in a sufficiently large random universe, there exist
low entropy regions that resemble our own universe
and/or brain (observer) \cite[Sec.3.8]{Barrow:86}.

\paradot{(A) All-a-Carte models}
The existence of true randomness is controversial and complicates
many considerations. So ToE (R) may be rejected on this ground,
but there is a simple deterministic computable variant.
Glue the natural numbers written in binary format,
1,10,11,100,101,110,111,1000,1001,... to one long string.
\beqn
  1 10 11 100 101 110 111 1000 1001 ...
\eeqn
The decimal version is known as Champernowne's number. Obviously it
contains every finite substring by construction. Indeed, it is a
Normal Number in the sense that it contains every substring of length $n$
with the same relative frequency ($2^{-n}$). Many irrational numbers
like $\sqrt{2}$, $\pi$, and $\e$ are conjectured to be normal. So
Champernowne's number and probably even $\sqrt{2}$ are perfect ToEs.

\paradot{Remarks}
I presume that every reader of this section at some point regarded
the remainder as bogus.
In a sense this paper is about a rational criterion to
decide whether a model is sane or insane.
The problem is that the line of sanity differs for different people
and different historical times.

Moving the earth out of the center of the universe was (and for some
even still is) insane.
The standard model is accepted by nearly all physicists as the
closest approximation to a ToE so far. Only outside physics, often
by opponents of reductionism, this view has been criticized. Some
respectable researchers including Nobel Laureates go further and
take string theory and even some Multiverse theories serious.
Universal ToE also has a few serious proponents. Whether Boltzmann's random
noise or my All-a-Carte ToE find adherers needs to be seen. For me,
Universal ToE (U) is the sanity critical point.
Indeed UToE will be investigated in greater detail in later sections.

References to the dogmatic Bible, Popper's misguided falsifiability
principle \cite{Stove:82,Gardner:01}, and wrong applications of
Ockham's razor are the most popular pseudo justifications of what
theories are (in)sane. More serious arguments involving the
usefulness of a theory will be discussed in the next section.

\section{Predictive Power \& Observer Localization}\label{secPPOL}

In the last section I have enumerated some models of (parts or
including) the universe, roughly sorted in increasing size of the
universe. Here I discuss their relative merits, in particular
their predictive power (precision and coverage).
Analytical or computational tractability also influences the
usefulness of a theory, but can be ignored when evaluating its
status as a ToE. For example, QED is computationally a nightmare, but
this does not at all affect its status as the theory of all
electrical, magnetic, and chemical processes.
On the other hand, we will see that localizing the observer, which
is usually not regarded as an issue, can be very important. The
latter has nothing to do with the quantum-mechanical measuring
process, although there may be some deeper yet to be explored
connection.

\paradot{Particle physics}
The standard model has more power and hence is closer to a ToE than
all effective theories (E) together.
String theory plus the right choice of compactification reduces to
the standard model, so has the same or superior power. The key
point here is the inclusion of the ``right choice of compactification''.
Without it, string theory is in some respect less powerful than SM.

\paradot{Baby universes}
Let us now turn to the cosmological models, in particular Smolin's
baby universe theory, in which infinitely many universes with
different properties exist. The theory ``explains'' why a universe
with our properties exist (since it includes universes with all
kinds of properties), but it has little predictive power. The baby
universe theory {\em plus} a specification in which universe we
happen to live would determine the value of the inter-universe
variables for our universe, and hence have much more predictive
power. So localizing ourselves increases the predictive power of the
theory.

\paradot{Universal ToE}
Let us consider the even larger universal multiverse. Assuming our
universe is computable, the multiverse generated by UToE contains
and hence perfectly describes our universe. But this is of little
use, since we can't use UToE for prediction. If we knew our
``position'' in this multiverse, we would know in which
(sub)universe we are. This is equivalent to knowing the program that
generates {\em our} universe. This program may be close to any of
the conventional cosmological models, which indeed have a lot of
predictive power. Since locating ourselves in UToE is equivalent and
hence as hard as finding a conventional ToE of our universe, we have
not gained much.

\paranodot{All-a-Carte models}
also contain and hence perfectly describe our universe. If and
only if we can localize ourselves, we can actually use it for
predictions. (For instance, if we knew we were in the center of
universe 001011011 we could predict that we will `see' 0010 when
`looking' to the left and 1011 when looking to the right.) Let $u$
be a snapshot of our space-time universe; a truly gargantuan string.
Locating ourselves means to (at least) locate $u$ in the multiverse.
We know that $u$ is the $u$'s number in Champernowne's sequence
(interpreting $u$ as a binary number), hence locating $u$ is
equivalent to specifying $u$. So a ToE based on normal numbers is
only useful if accompanied by the gargantuan snapshot $u$ of our
universe. In light of this, an ``All-a-Carte'' ToE (without knowing $u$)
is rather a theory of nothing than a theory of everything.

\paradot{Localization within our universe}
The loss of predictive power when enlarging a universe to a
multiverse model has nothing to do with multiverses per se. Indeed,
the distinction between a universe and a multiverse is not absolute.
For instance, Champernowne's number could also be interpreted as a
single universe, rather than a multiverse. It could be regarded as
an extreme form of the infinite fantasia land from the NeverEnding
Story, where everything happens somewhere. Champernowne's number
constitutes a perfect map of the All-a-Carte universe, but the map is
useless unless you know where you are.
Similarly but less extreme, the inflation model produces a universe
that is vastly larger than its visible part, and different regions
may have different properties.

\paradot{Egocentric to Geocentric model}
Consider now the ``small'' scale of our daily life. A young
child believes it is the center of the world. 
Localization is trivial. It is always at ``coordinate'' (0,0,0).
Later it learns that it is just one among a few billion other people
and as little or much special as any other person thinks of
themself. In a sense we replace our egocentric coordinate system by
one with origin (0,0,0) in the center of Earth. The move away from
an egocentric world view has many social advantages, but dis-answers
one question: Why am I this particular person and not any other? (It
also comes at the cost of constantly having to balance egoistic with
altruistic behavior.)

\paradot{Geocentric to Heliocentric model}
While being expelled from the center of the world as an individual,
in the geocentric model, at least the human race as a whole remains
in the center of the world, with the remaining (dead?)
universe revolving around {\em us}.
The heliocentric model puts Sun at (0,0,0) and degrades Earth to
planet number 3 out of 8. The astronomic advantages are clear, but
dis-answers one question: Why this planet and not one of the others?
Typically we are muzzled by questionable anthropic arguments
\cite{Bostrom:02,Smolin:04}. (Another scientific cost is the
necessity now to switch between coordinate systems, since the ego-
and geocentric views are still useful.)

\paradot{Heliocentric to modern cosmological model}
The next coup of astronomers was to degrade our Sun to one star
among billions of stars in our milky way, and our milky way to one
galaxy out of billions of others. It is generally accepted that the
question why we are in this particular galaxy in this particular
solar system is essentially unanswerable.

\paradot{Summary}
The exemplary discussion above has hopefully convinced the reader
that we indeed lose something (some predictive power) when
progressing to too large universe and multiverse models.
Historically, the higher predictive power of the large-universe
models (in which we are seemingly randomly placed) overshadowed the
few extra questions they raised compared to the smaller
ego/geo/helio-centric models. (we're not concerned here with the
psychological disadvantages/damage, which may be large).
But the discussion of the (physical, universal, random, and
all-a-carte) multiverse theories has shown that pushing this
progression too far will at some point harm predictive power. We saw
that this has to do with the increasing difficulty to localize the
observer.

\section{Complete ToEs (CToEs)}\label{secCToEs}

A ToE by definition is a perfect model of the universe. It should
allow to predict all phenomena. Most ToEs require a specification of
some initial conditions, e.g.\ the state at the big bang, and how
the state evolves in time (the equations of motion). In general, a
ToE is a program that in principle can ``simulate'' the whole
universe.
An All-a-Carte universe perfectly satisfies this condition but
apparently is rather a theory of nothing than a theory of
everything. So meeting the simulation condition is not sufficient
for qualifying as a Complete ToE.
We have seen that (objective) ToEs can be completed by specifying
the location of the observer. This allows us to make useful
predictions from our (subjective) viewpoint. We call a ToE
plus observer localization a subjective or complete ToE.
If we allow for stochastic (quantum) universes we also need to include the noise.
If we consider (human) observers with limited perception ability we
need to take that into account too. So

\paranodot{A complete ToE needs specification of}
\begin{itemize}\parskip=0ex\parsep=0ex\itemsep=0ex
\item[(i)] initial conditions
\item[(e)] state evolution
\item[(l)] localization of observer
\item[(n)] random noise
\item[(o)] perception ability of observer
\end{itemize}
We will ignore noise and perception ability in the following and
resume to these issues in Sections \ref{secExt} and \ref{secCToE},
respectively.
Next we need a way to compare ToEs.

\paradot{Epistemology}
I assume that the observers' experience of the world consists of a
single temporal binary sequence which gets longer with time.
This is definitely true if the observer is a robot equipped with
sensors like a video camera whose signal is converted to a digital
data stream, fed into a digital computer and stored in a binary file
of increasing length. In humans, the signal transmitted by the optic
and other sensory nerves could play the role of the digital data
stream.
Of course (most) human observers do not possess photographic memory.
We can deal with this limitation in various ways: digitally record
and make accessible upon request the nerve signals from birth till
now, or allow for uncertain or partially remembered observations.
Classical philosophical theories of knowledge \cite{Alchin:06}
(e.g.\ as justified true belief) operate on a much higher conceptual
level and therefore require stronger (and hence more disputable)
philosophical presuppositions. In my minimalist ``spartan''
information-theoretic epistemology, a bit-string is the only
observation, and all higher ontologies are constructed from it
and are pure ``imagination''.

\paradot{Predictive power and elegance}
Whatever the intermediary guiding principles for designing
theories/models (elegance, symmetries, tractability, consistency),
the ultimate judge is predictive success.
Unfortunately we can never be sure whether a given ToE makes correct
predictions in the future. After all we cannot rule out that the
world suddenly changes tomorrow in a totally unexpected way (cf.\
the quote at beginning of this article). We have to compare theories
based on their predictive success in the past. It is also clear that
the latter is not enough: For every model we can construct an
alternative model that behaves identically in the past but makes
different predictions from, say, year 2020 on. Popper's
falsifiability dogma is little helpful. Beyond postdictive success,
the guiding principle in designing and selecting theories,
especially in physics, is elegance and mathematical consistency. The
predictive power of the first heliocentric model was not superior to
the geocentric one, but it was much simpler. In more profane terms,
it has significantly less parameters that need to be specified.

\paranodot{Ockham's razor}
suitably interpreted tells us to choose the
simpler among two or more otherwise equally good theories.
For justifications of Ockham's razor, see \cite{Li:08} and Section \ref{secPOR}.
Some even argue that by definition, science is about applying Ockham's razor, see
\cite{Hutter:04uaibook}.
For a discussion in the context of theories in physics, see \cite{Gellmann:94}.
It is beyond the scope of this paper to repeat these considerations.
In Sections \ref{secCToEs} and \ref{secPOR} I will show that simpler theories
more likely lead to correct predictions, and therefore Ockham's
razor is suitable for finding ToEs.

\paradot{Complexity of a ToE}
In order to apply Ockham's razor in a non-heuristic way, we need to
quantify simplicity or complexity. Roughly, the complexity of a
theory can be defined as the number of symbols one needs to write
the theory down. More precisely, write down a program for the state
evolution together with the initial conditions, and define the
complexity of the theory as the size in bits of the file that contains
the program.
This quantification is known as algorithmic information or
Kolmogorov complexity \cite{Li:08} and is consistent with our intuition,
since an elegant theory will have a shorter program than an
inelegant one, and extra parameters need extra space to code, resulting in
longer programs \cite{Wallace:05,Gruenwald:07book}.
From now on I identify theories with programs and write $\Length(q)$
for the length=complexity of program=theory $q$.

\paradot{Standard model versus string theory}
To keep the discussion simple, let us pretend that standard model
(SM) + gravity (G) and string theory (S) both qualify as ToEs.
SM+Gravity is a mixture of a few relatively elegant theories, but
contains about 20 parameters that need to be specified. String
theory is truly elegant, but ensuring that it reduces to the
standard model needs sophisticated extra assumptions (e.g. the right
compactification).

SM+G can be written down in one line, plus we have to give 20+
constants, so lets say one page. The meaning (the axioms) of all
symbols and operators require another page. Then we need the
basics, natural, real, complex numbers, sets (ZFC), etc., which is
another page. That makes 3 pages for a complete description in
first-order logic. There are a lot of subtleties though: (a) The
axioms are likely mathematically inconsistent, (b) it's not
immediately clear how the axioms lead to a program simulating our
universe, (c) the theory does not predict the outcome of random
events, and (d) some other problems. So to transform the description
into a C program simulating our universe, needs a couple of pages
more, but I would estimate around 10 pages overall suffices, which
is about 20'000 symbols=bytes. Of course this program will be (i) a very
inefficient simulation and (ii) a very naive coding of SM+G. I
conjecture that the {\em shortest} program for SM+G on a universal
Turing machine is much shorter, maybe even only one tenth of this.
The numbers are only a quick rule-of-thumb guess. If we start from
string theory (S), we need about the same length. S is {\em much}
more elegant, but we need to code the compactification to describe
our universe, which effectively amounts to the same. Note that
everything else in the world (all other physics, chemistry, etc,) is
emergent.

It would require a major effort to quantify which theory is the
simpler one in the sense defined above, but I think it would be
worth the effort. It is a quantitative objective way to decide
between theories that are (so far) predictively indistinguishable.

\paradot{CToE selection principle}
It is trivial to write down a program for an All-a-Carte multiverse
(A). It is also not too hard to write a program for the universal
multiverse (U), see Section \ref{secUToE}. Lengthwise (A) easily
wins over (U), and (U) easily wins over (P) and (S), but as
discussed (A) and (U) have serious defects. On the other hand, these
theories can only be used for predictions after extra
specifications: Roughly, for (A) this amounts to tabling the whole
universe, (U) requires defining a ToE in the conventional sense, (P)
needs 20 or so parameters and (S) a compactification scheme. Hence
localization-wise (P) and (S) easily win over (U), and (U) easily
wins over (A).
Given this trade-off, it now nearly suggests 
itself that we should include the description length of the observer
location in our ToE evaluation measure. That is,
\beqn
  \mbox{among two CToEs, select the one that has shorter overall length}
\eeqn
\beq\label{orctoe}
  \Length(i)+\Length(e)+\Length(l)
\eeq
For an All-a-Carte multiverse, the last term contains the gargantuan
string $u$, catapulting it from the shortest ToE to the longest
CToE, hence (A) will not minimize \req{orctoe}.

\paradot{ToE versus UToE}
Consider any (C)ToE and its program $q$, e.g.\ (P) or (S). Since (U) runs
all programs including $q$, specifying $q$ means localizing (C)ToE
$q$ in (U). So (U)+$q$ is a CToE whose length is just some constant
bits (the simulation part of (U)) more than that of (C)ToE $q$. So
whatever (C)ToE physicists come up with, (U) is nearly as good as
this theory. This essentially clarifies the paradoxical status of
(U). Naked, (U) is a theory of nothing, but in combination with
another ToE it excels to a good CToE, albeit slightly longer=worse
than the latter.

\paradot{Localization within our universe}
So far we have only localized our universe in the multiverse, but
not ourselves in the universe.
To localize our Sun, we could e.g.\ sort (and index) stars by their
creation date, which the model (i)+(e) provides. Most stars last for
1-10 billion years (say an average of 5 billion years). The universe
is 14 billion years old, so most stars may be 3rd generation (Sun
definitely is), so the total number of stars that have ever existed
should very roughly be 3 times the current number of stars of about
$10^{11}\times 10^{11}$. Probably ``3'' is very crude, but this
doesn't really matter for sake of the argument.
In order to localize our Sun we only need its index, which can be
coded in about $\lb (3\times 10^{11}\times 10^{11})\doteq 75$ bits.
Similarly we can sort and index planets and observers. To localize
earth among the 8 planets needs 3 bits. To localize yourself among 7
billion humans needs 33 bits.
Alternatively one could simply specify the $(x,y,z,t)$ coordinate of
the observer, which requires more but still only very few bits.
These localization penalties are tiny compared to the difference in
predictive power (to be quantified later) of the various theories
(ego/geo/helio/cosmo). This explains and justifies theories of large
universes in which we occupy a random location.

\section{Complete ToE - Formalization}\label{secCToE}

This section formalizes the CToE selection principle and what
accounts for a CToE. Universal Turing machines are used to formalize
the notion of programs as models for generating our universe and our
observations. I also introduce more realistic observers with limited
perception ability.

\paradot{Objective ToE}
Since we essentially identify a ToE with a program generating a
universe, we need to fix some general purpose programming language
on a general purpose computer. In theoretical computer science, the
standard model is a so-called Universal Turing Machine ($\UTM$)
\cite{Li:08}. It takes a program coded as a finite binary string
$q\in\{0,1\}^*$, executes it and outputs a finite or infinite binary
string $u\in\{0,1\}^*\cup\{0,1\}^\infty$. The details do not matter
to us, since drawn conclusions are typically independent of them. In
this section we only consider $q$ with infinite output
\beqn
  \UTM(q)=u_1^q u_2^q u_3^q\,... =:u_{1:\infty}^q
\eeqn
In our case, $u_{1:\infty}^q$ will be the space-time universe (or
multiverse) generated by ToE candidate $q$. So $q$ incorporates
items (i) and (e) of Section \ref{secCToEs}. Surely our universe
doesn't look like a bit string, but can be coded as one as explained
in Sections \ref{secSToE} and \ref{secExt}. We have some simple
coding in mind, e.g.\ $u_{1:N}^q$ being the (fictitious) binary data
file of a high-resolution 3D movie of the whole universe from big
bang to big crunch, augmented by $u_{N+1:\infty}^q\equiv 0$ if the
universe is finite. Again, the details do not matter.

\paradot{Observational process and subjective complete ToE}
As we have demonstrated it is also important to localize the
observer. In order to avoid potential qualms with modeling human
observers, consider as a surrogate a (conventional not extra cosmic) video
camera filming=observing parts of the world. The camera may be fixed
on Earth or installed on an autonomous robot. It records part of the
universe $u$ denoted by $o=o_{1:\infty}$. (If the lifetime of the
observer is finite, we append zeros to the finite observation
$o_{1:N}$).

I only consider {\em direct} observations like with a camera.
Electrons or atomic decays or quasars are not directly observed, but
with some (classical) instrument. It is the indicator or camera image of
the instrument that is observed (which physicists then usually interpret).
This setup avoids having to deal with
any form of informal correspondence between theory and real world,
or with subtleties of the quantum-mechanical measurement process.
The only philosophical presupposition I make is
that it is possible to determine uncontroversially whether two finite binary
strings (on paper or file) are the same or differ in some bits.

In a computable universe, the observational process within it, is
obviously also computable, i.e.\ there exists a program
$s\in\{0,1\}^*$ that extracts observations $o$ from universe $u$.
Formally
\beq
  \UTM(s,u_{1:\infty}^q) = o_{1:\infty}^{sq}
\eeq
where we have extended the definition of $\UTM$ to allow access to
an extra infinite input stream $u_{1:\infty}^q$. So
$o_{1:\infty}^{sq}$ is the sequence observed by subject $s$
in universe $u_{1:\infty}^q$ generated by $q$.
Program $s$ contains
all information about the location and orientation and perception
abilities of the observer/camera, hence specifies not only item (l)
but also item (o) of Section \ref{secCToEs}.

A Complete ToE (CToE) consists of a specification of a (ToE,subject)
pair $(q,s)$. Since it includes $s$ it is a Subjective ToE.

\paradot{CToE selection principle}
So far, $s$ and $q$ were fictitious subjects and universe programs.
Let $o_{1:t}^{true}$ be the past observations of some concrete
observer in our universe, e.g.\ your own personal experience of the
world from birth till today. The future observations
$o_{t+1:\infty}^{true}$ are of course unknown. By definition,
$o_{1:t}$ contains {\em all} available experience of the observer,
including e.g.\ outcomes of scientific experiments, school
education, read books, etc.

The observation sequence $o_{1:\infty}^{sq}$ generated by a
correct CToE must be consistent with the true observations
$o_{1:t}^{true}$. If $o_{1:t}^{sq}$ would differ from $o_{1:t}^{true}$
(in a single bit) the subject would have `experimental' evidence
that $(q,s)$ is not a perfect CToE.
We can now formalize the CToE selection principle as follows
\beqn
  \mbox{Among a given set of perfect ($o_{1:t}^{sq}=o_{1:t}^{true}$) CToEs
  $\{(q,s)\}$}
\eeqn
\beq\label{Lqs}
  \mbox{select the one of smallest length }  \Length(q)+\Length(s)
\eeq
Minimizing length is motivated by Ockham's razor.
Inclusion of $s$ is necessary to avoid degenerate ToEs like (U) and
(A).
The selected CToE $(q^*,s^*)$ can and should then be used for
forecasting future observations via
$...o_{t+1:\infty}^{forecast}=\UTM(s^*,u_{1:\infty}^{q^*})$.

\section{Universal ToE - Formalization}\label{secUToE}

The Universal ToE is a sanity critical point in the development of
ToEs, and will formally be defined and investigated in this section.

\paradot{Definition of Universal ToE}
The Universal ToE generates all computable universes.
The generated multiverse can be depicted as an infinite
matrix in which each row corresponds to one universe.
\beqn\arraycolsep0ex
  \begin{array}{c|cccccc}
    q\; & \multicolumn{3}{c}{\UTM(q)}  &  \\ \hline
 \epstr & u_1^\epstr & u_2^\epstr & u_3^\epstr & u_4^\epstr & u_5^\epstr & \cdots  \\
    0   & u_1^0      & u_2^0      & u_3^0      & u_4^0      & \cdots     & \cdots  \\
    1   & u_1^1      & u_2^1      & u_3^1      & \cdots     & \cdots     &         \\
   00\; & u_1^{00}   & u_2^{00}   & \cdots     & \cdots     &            &         \\
 \vdots & \vdots     & \vdots     & \vdots     &  \\
  \end{array}
\eeqn
To fit this into our framework we need to define a single program
$\breve q$ that generates a single string corresponding to this matrix.
The
standard way to linearize an infinite matrix is to dovetail in
diagonal serpentines though the matrix:
\beqn
  \breve u_{1:\infty} := u_1^\epstr u_1^0 u_2^\epstr u_3^\epstr
  u_2^0 u_1^1 u_1^{00} u_2^1 u_3^0 u_4^\epstr u_5^\epstr
  u_4^0 u_3^1 u_2^{00} ...
\eeqn
Formally, define a bijection
$i=\langle q,k\rangle$ between a (program, location) pair $(q,k)$
and the natural numbers $\SetN\ni i$, and define $\breve u_i:=u_k^q$.
It is not hard to construct an explicit program $\breve q$ for
$\UTM$ that computes $\breve u_{1:\infty}=u_{1:\infty}^{\breve
q}=\UTM(\breve q)$.

One might think that it would have been simpler or more natural to
generalize Turing machines to have matrix ``tapes''. But this is
deceiving. If we allow for Turing machines with matrix output, we
also should allow for and enumerate all programs $q$ that have a
matrix output. This leads to a 3d tensor that needs to be converted
to a 2d matrix, which is no simpler than the linearization above.

\paradot{Partial ToEs}
Cutting the universes into bits and interweaving them into one string
might appear messy, but is unproblematic for two reasons: First, the
bijection $i=\langle q,k\rangle$ is very simple, so any particular
universe string $u^q$ can easily be recovered from $\breve u$.
Second, such an extraction will be included in the localization /
observational process $s$, i.e.\ $s$ will contain a specification of
the relevant universe $q$ and which bits $k$ are to be observed.

More problematic is that many $q$ will not produce an infinite
universe. This can be fixed as follows: First, we need to be more
precise about what it means for $\UTM(q)$ to write $u^q$. We
introduce an extra symbol `\#' for `undefined' and set each bit
$u_i^q$ initially to `\#'. The $\UTM$ running $q$ can output bits
in any order but can overwrite each location \#
only once, either with a 0 or with a 1. We implicitly assumed
this model above, and similarly for $s$. Now we (have to) also allow
for $q$ that leave some or all bits unspecified.

The interleaving computation $\UTM(s,\UTM(q)) = o$ of $s$ and $q$
works as follows: Whenever $s$ wants to read a bit from
$u_{1:\infty}^q$ that $q$ has not (yet) written, control is
transferred to $q$ until this bit is written. If it is never
written, then $o$ will be only partially defined, but such $s$ are
usually not considered. (If the undefined location is before $t$,
CToE $(q,s)$ is not perfect, since $o_{1:t}^{true}$ is completely defined.)

Alternatively one may define a more complex dynamic bijection
$\langle\cdot,\cdot\rangle$ that orders the bits in the order they
are created, or one resorts to generalized Turing machines
\cite{Schmidhuber:00toe,Schmidhuber:02gtm} which can overwrite
locations and also greatly increase the set of describable
universes. These variants (allow to) make $\breve u$ and all $u^q$
complete (no `\#' symbol, tape$\,\in\{0,1\}^\infty$).

\paradot{ToE versus UToE}
We can formalize the argument in the last section of simulating a ToE
by UToE as follows: If $(q,s)$ is a CToE, then $(\breve q,\tilde s)$
based on UToE $\breve q$ and observer $\tilde s:=rqs$, where program
$r$ extracts $u^q$ from $\breve u$ and then $o^{sq}$ from $u^q$, is
an equivalent but slightly larger CToE, since $\UTM(\tilde s,\breve
u)=o^{qs}=\UTM(s,u^q)$ by definition of $\tilde s$ and
$\Length(\breve q)+\Length(\tilde s) = \Length(q)+\Length(s)+O(1)$.

\paradot{The best CToE}
Finally, one may define the best CToE (of an observer with experience
$o_{1:t}^{true}$) as
\beqn
  \mbox{UCToE} := \arg\min_{q,s}\{\Length(q)+\Length(s) :
  o_{1:t}^{sq}=o_{1:t}^{true} \}
\eeqn
where $o_{1:\infty}^{sq} = \UTM(s,\UTM(q))$. This may be regarded as
a formalization of the holy grail in physics; of finding such a TOE.

\section{Extensions}\label{secExt}

Our CToE selection principle is applicable to perfect,
deterministic, discrete, and complete models $q$ of our universe.
None of the existing sane world models is of this kind. In this
section I extend the CToE selection principle to more realistic,
partial, approximate, probabilistic, and/or parametric models for finite,
infinite and even continuous universes.

\paradot{Partial theories}
Not all interesting theories are ToEs. Indeed, most theories are
only partial models of aspects of our world.

We can reduce the problem of selecting good partial theories to CToE
selection as follows: Let $o_{1:t}^{true}$ be the complete
observation, and $(q,s)$ be some theory explaining only some
observations but not all. For instance, $q$ might be the
heliocentric model and $s$ be such that all bits in $o_{1:t}^{true}$
that correspond to planetary positions are predicted correctly. The
other bits in $o_{1:t}^{qs}$ are undefined, e.g.\ the position of
cars. We can augment $q$ with a (huge) table $b$ of all bits for
which $o_i^{qs}\neq o_i^{true}$. Together, $(q,b,s)$ allows to
reconstruct $o_{1:t}^{true}$ exactly. Hence for two different
theories, the one with smaller length
\beq\label{Lqbs}
  \Length(q) + \Length(b) + \Length(s)
\eeq
should be selected. We can actually spare ourselves from tabling all
those bits that are unpredicted by all $q$ under consideration,
since they contribute the same overall constant. So when comparing
two theories it is sufficient to consider only those observations
that are correctly predicted by one (or both) theories.

If two partial theories $(q,s)$ and $(q',s')$
predict the same phenomena equally well (i.e.\
$o_{1:t}^{qs}=o_{1:t}^{q's'}\neq o_{1:t}^{true}$), then $b=b'$
and minimizing \req{Lqbs} reduces to minimizing \req{Lqs}.

\paradot{Approximate theories}
Most theories are not perfect but only approximate reality, even in
their limited domain. The geocentric model is less accurate than the
heliocentric model, Newton's mechanics approximates general
relativity, etc. Approximate theories can be viewed as a version of
partial theories. For example, consider predicting locations of
planets with locations being coded by (truncated) real numbers in
binary representation, then Einstein gets more bits right than
Newton. The remaining erroneous bits could be tabled as above.
Errors are often more subtle than simple bit errors, in which case
correction programs rather than just tables are needed.

\paradot{Celestial example}
The ancient celestial models just capture the movement of some
celestial bodies, and even those only imperfectly. Nevertheless it
is interesting to compare them. Let us take as our corpus of
observations $o_{1:t}^{true}$, say, all astronomical tables
available in the year 1600, and ignore all other experience.

The geocentric model $q^G$ more or less directly describes the
observations, hence $s^G$ is relatively simple. In the heliocentric
model $q^H$ it is necessary to include in $s^H$ a non-trivial
coordinate transformation to explain the geocentric astronomical
data. Assuming both models were perfect, then, if and only if $q^H$
is simpler than $q^G$ by a margin that is larger than the extra
complications due to the coordinate transformation
($\Length(q^G)-\Length(q^H) > \Length(s^H)-\Length(s^G)$), we should
regard the heliocentric model as better.

If/since the heliocentric model is more accurate, we have to
additionally penalize the geocentric model by the number of bits it
doesn't predict correctly. This clearly makes the heliocentric model
superior.

\paradot{Probabilistic theories}
Contrary to a deterministic theory that predicts the future from the
past for sure, a probabilistic theory assigns to each future a
certain chance that it will occur. Equivalently, a deterministic
universe is described by some string $u$, while a probabilistic
universe is described by some probability distribution $Q(u)$, the
a priori probability of $u$. (In the special case of $Q(u')=1$ for
$u'=u$ and 0 else, $Q$ describes the deterministic universe $u$.)
Similarly, the observational process may be probabilistic. Let
$S(o|u)$ be the probability of observing $o$ in universe $u$.
Together, $(Q,S)$ is a probabilistic CToE that predicts observation
$o$ with probability $P(o)=\sum_u S(o|u)Q(u)$. A computable
probabilistic CToE is one for which there exist programs (of
lengths $\Length(Q)$ and $\Length(S)$) that compute the functions
$Q(\cdot)$ and $S(\cdot|\cdot)$.

Consider now the true observation $o_{1:t}^{true}$. The larger
$P(o_{1:t}^{true})$ the ``better'' is $(Q,S)$. In the degenerate
deterministic case, $P(o_{1:t}^{true})=1$ is maximal for a correct
CToE, and 0 for a wrong one. In every other case, $(Q,S)$ is only a
partial theory that needs completion, since it does not
compute $o_{1:t}^{true}$. Given $P$, it is possible to code
$o_{1:t}^{true}$ in $|\lb P(o_{1:t}^{true})|$ bits (arithmetic or
Shannon-Fano code). Assuming that $o_{1:t}^{true}$ is indeed sampled
from $P$, one can show that with high probability this is the
shortest possible code. So there exists an effective description
of $o_{1:t}^{true}$ of length
\beq\label{QSPcode}
  \Length(Q) + \Length(S) + |\lb P(o_{1:t}^{true})|
\eeq
This expression should be used (minimized) when comparing
probabilistic CToEs. The principle is reminiscent of classical
two-part Minimum Encoding Length principles like MML and MDL
\cite{Wallace:05,Gruenwald:07book}.
Note that the noise corresponds to the errors and the log term to the
error table of the previous paragraphs.

\paradot{Probabilistic examples}
Assume $S(o|o)=1\,\forall o$ and consider the observation sequence
$o_{1:t}^{true} = u_{1:t}^{true} =
1100\-1001\-0000\-1111\-1101\-1010\-10100$. If we assume this is a
sequence of fair coin flips, then $Q(o_{1:t})=P(o_{1:t})=2^{-t}$ are
very simple functions, but $|\lb P(o_{1:t})|=t$ is large. If we assume that
$o_{1:t}^{true}$ is the binary expansion of $\pi$ (which it is),
then the corresponding deterministic $Q$ is somewhat more complex,
but $|\lb P(o_{1:t}^{true})|=0$. So for sufficiently large $t$, the
deterministic model of $\pi$ is selected, since it leads to a shorter code
\req{QSPcode} than the fair-coin-flip model.

Quantum theory is (argued by physicists to be) truly random. Hence
all modern ToE candidates (P+G, S, C, M) are probabilistic. This
yields huge additive constants $|\lb P(o_{1:t}^{true})|$ to the
otherwise quite elegant theories $Q$. Schmidhuber
\cite{Schmidhuber:97brauer,Schmidhuber:00toe} argues that all
apparent physical randomness is actually only pseudo random, i.e.\
generated by a small program. If this is true and we could find the
random number generator, we could instantly predict all apparent
quantum-mechanical random events. This would be a true improvement
of existing theories, and indeed the corresponding CToE would be
significantly shorter.
In \cite[Sec.8.6.2]{Hutter:04uaibook} I give an argument why believing in true
random noise may be an unscientific position.

\paradot{Theories with parameters}
Many theories in physics depend on real-valued parameters.
Since observations have finite accuracy,
it is sufficient to specify these parameters to some finite accuracy.
Hence the theories including their finite-precision parameters
can be coded in finite length.
There are general results and techniques \cite{Wallace:05,Gruenwald:07book}
that allow a comfortable handling of all this. For instance,
for smooth parametric models, a parameter accuracy of $O(1/\sqrt{n})$ is
needed, which requires $\fr12\lb n+O(1)$ bits per parameter.
The explicable $O(1)$ term depends on the smoothness of the model
and prevents `cheating' (e.g.\ zipping two parameters into one).

\paradot{Infinite and continuous universes}
So far we have assumed that each time-slice through our universe can
be described in finitely many bits and time is discrete. Assume our
universe were the infinite continuous 3+1 dimensional Minkowski
space $\SetR^4$ occupied by (tiny) balls (``particles''). Consider
all points $(x,y,z,t)\in\SetR^4$ with rational coordinates, and let
$i=\langle x,y,z,t\rangle$ be a bijection to the natural numbers
similarly to the dovetailing in Section \ref{secUToE}. Let $u_i=1$ if
$(x,y,z,t)$ is occupied by a particle and 0 otherwise. String
$u_{1:\infty}$ is an exact description of this universe. The above
idea generalizes to any so-called separable mathematical space. Since all
spaces occurring in established physical theories are separable,
there is currently no ToE candidate that requires uncountable
universes. Maybe continuous theories are just convenient
approximations of deeper discrete theories.
An even more fundamental argument put forward in this context by
\cite{Schmidhuber:00toe} is that the Loewenheim-Skolem
theorem (an apparent paradox) implies that Zermelo-Fraenkel set
theory (ZFC) has a countably infinite model. Since all physical
theories so far are formalizable in ZFC, it follows they all have a
countable model. For some strange reason (possibly an historical
artifact), the adopted uncountable interpretation seems just more
convenient.

\paradot{Multiple theories}
Some proponents of pluralism and some opponents of reductionism
argue that we need multiple theories on multiple
scales for different (overlapping) application domains.
They argue that a ToE is not desirable and/or not possible.
Here I give a reason why we {\em need} one {\em single} fundamental
theory (with all other theories having to be regarded as
approximations):
Consider two Theories ($T1$ and $T2$) with (proclaimed) application
domains $A1$ and $A2$, respectively.

If predictions of $T1$ and $T2$ coincide on their intersection
$A1\cap A2$ (or if $A1$ and $A2$ are disjoint), we can trivially
``unify'' $T1$ and $T2$ to one theory $T$ by taking their union. Of
course, if this does not result in any simplification, i.e.\ if
$\Length(T)=\Length(T1)+\Length(T2)$, we gain nothing. But
since nearly all modern theories have some common basis, e.g.\ use
natural or real numbers, a formal unification of the generating
programs nearly always leads to
$\Length(q)<\Length(q_1)+\Length(q_2)$.

The interesting case is when $T1$ and $T2$ lead to different
forecasts on $A1\cap A2$.
For instance, particle versus wave theory with the atomic world at
their intersection, unified by quantum theory.
Then we need a reconciliation of $T1$ and $T2$, that is, a single
theory $T$ for $A1\cup A2$. Ockham's razor tells us to choose a
simple (elegant) unification. This rules out naive/ugly/complex
solutions like developing a third theory for $A1\cap A2$ or
attributing parts of $A1\cap A2$ to $T1$ or $T2$ as one sees fit, or
averaging the predictions of $T1$ and $T2$. Of course $T$ must be
consistent with the observations.

Pluralism on a meta level, i.e.\ allowing besides Ockham's razor
other principles for selecting theories, has the same problem on a
meta-level: which principle should one use in a concrete situation?
To argue that this (or any other) problem cannot be
formalized/quantized/mechanized would be (close to) an
anti-scientific attitude.

\section{Justification of Ockham's Razor}\label{secPOR}

We now prove Ockham's razor under the assumptions stated below and
compare it to the No Free Lunch myth.
The result itself is not novel \cite{Schmidhuber:00toe}. The
intention and contribution is to provide an elementary but still
sufficiently formal argument, which in particular is free of more
sophisticated concepts like Solomonoff's a-priori distribution.

Ockham's razor principle demands to ``take the simplest theory
consistent with the observations''.

\begin{quote}\it
Ockham's razor could be regarded as correct if among all considered
theories, the one selected by Ockham's razor is the one that most
likely leads to correct predictions.
\end{quote}

\paradot{Assumptions}
Assume we live in the universal multiverse $\breve u$ that consists
of all computable universes, i.e.\ UToE is a correct/true/perfect
ToE. Since every computable universe is contained in UToE, it is at
least under the computability assumption impossible to disprove this
assumptions.
The second assumption we make is that our location in the multiverse
is random. We can divide this into two steps: First, the universe
$u^q$ in which we happen to be is chosen randomly. Second, our
``location'' $s$ within $u^q$ is chosen at random.
We call these the {\em universal self-sampling assumption}. The crucial
difference to the informal anthropic self-sampling assumption used
in doomsday arguments is discussed below.

Recall the observer program $\tilde s:=rqs$ introduced in Section
\ref{secCToE}. We will make the simplifying assumption that $s$ is the
identity, i.e.\ restrict ourselves to ``objective'' observers that
observe their universe completely: $\UTM(\tilde s,\breve
u)=o^{qs}=\UTM(s,u^q)=u^q=\UTM(q)$. Formally, the universal
self-sampling assumption can be stated as follows:
\begin{quote}\it
  A priori it is equally likely to be in any of the universes
  $u^q$ generated by some program $q\in\{0,1\}^*$.
\end{quote}
To be precise, we consider all programs with length bounded by some
constant $L$, and take the limit $L\to\infty$.

\paradot{Counting consistent universes}
Let $o_{1:t}^{true}=u_{1:t}^{true}$ be the universe observed so far
and
\beqn
  Q_L:=\{q : \Length(q)\leq L \mbox{ and } \UTM(q)=u_{1:t}^{true}* \}
\eeqn
be the set of all consistent universes (which is non-empty for large
$L$), where * is any continuation of $u_{1:t}^{true}$. Given
$u_{1:t}^{true}$, we know we are in one of the universes in $Q_L$,
which implies by the universal self-sampling assumption a uniform
sampling in $Q_L$. Let
\beqn
  q_{min} := \arg\min_q\{\Length(q):q\in Q_L\} \qmbox{and}
  l:=\Length(q_{min})
\eeqn
be the shortest consistent $q$ and its length, respectively.
Adding (unread) ``garbage'' $g$ after the end of a program $q$
does not change its behavior, i.e.\ if $q\in Q_L$, then also $qg\in
Q_L$ provided that $\Length(qg)\leq L$. Hence for every $g$ with
$\Length(g)\leq L-l$, we have $q_{min}g\in Q_L$. Since there are about
$2^{L-l}$ such $g$, we have $|Q_L|\gtrsim 2^{L-l}$. It is a deep
theorem in algorithmic information theory \cite{Li:08} that
there are also not significantly more than $2^{L-l}$ programs $q$
equivalent to $q_{min}$. The proof idea is as follows: One can show
that if there are many long equivalent programs, then there must
also be a short one. In our case the shortest one is $q_{min}$,
which upper bounds the number of long programs. Together this shows
that
\beqn
  |Q_L| \;\approx \; 2^{L-l}
\eeqn

\paradot{Probabilistic prediction}
Given observations $u_{1:t}^{true}$ we now determine the probability
of being in a universe that continues with $u_{t+1:n}$, where $n>t$.
Similarly to the previous paragraph we can approximately count the
number of such universes:
\bqan
  Q_L^n &:=& \{ q: \Length(q)\leq L \mbox{ and } \UTM(q)=u_{1:t}^{true}u_{t+1:n}*
  \} \;\subset Q_L
\\
  q_{min}^n &:=& \arg\min_q\{\Length(q):q\in Q_L^n\} \qmbox{and}
  l_n:=\Length(q_{min}^n)
\\
  |Q_L^n| &\approx& 2^{L-l_n}
\eqan
The probability of being in a universe with future $u_{t+1:n}$ given
$u_{1:t}^{true}$ is determined by their relative number
\beq\label{eqMKm}
  P(u_{t+1:n}|u_{1:t}^{true})
  \;=\; {|Q_L^n|\over|Q_L|}
  \;\approx\; 2^{-(l_n-l)}
\eeq
which is (asymptotically) independent of $L$.

\paradot{Ockham's razor}
Relation \req{eqMKm} implies that the most likely continuation $\hat
u_{t+1:n}:=\arg\max_{u_{t+1:n}}P(u_{t+1:n}|u_{1:t}^{true})$ is
(approximately) the one that minimizes $l_n$. By definition,
$q_{min}$ is the shortest program in $Q_L=\bigcup_{u_{t+1:n}}Q_L^n$.
Therefore
\beqn
  P(\hat u_{t+1:n}|u_{1:t}^{true}) \approx
  P(u_{t+1:n}^{q_{min}}|u_{1:t}^{true})
\eeqn
The accuracy of $\approx$ is clarified later. In words
\begin{quote}\it
We are most likely in a universe that is (equivalent to) the
simplest universe consistent with our past observations.
\end{quote}
This shows that Ockham's razor selects the theory that most likely
leads to correct predictions, and hence proves (under the stated
assumptions) that Ockham's razor is correct.
\begin{quote}\it
Ockham's razor is correct under the universal self-sampling assumption.
\end{quote}

\paradot{Discussion}
It is important to note that the universal self-sampling assumption
has not by itself any bias towards simple models $q$. Indeed, most
$q$ in $Q_L$ have length close to $L$, and since we sample uniformly
from $Q_L$ this actually represents a huge bias towards large
models for $L\to\infty$.

The result is also largely independent of the uniform sampling
assumption. For instance, sampling a length $l\in\SetN$ w.r.t.\ any
reasonable (i.e.\ slower than exponentially decreasing) distribution
and then $q$ of length $l$ uniformly leads to the same conclusion.

How reasonable is the UToE? We have already discussed that it is
nearly but not quite as good as any other correct ToE. The
philosophical, albeit not practical advantage of UToE is that it is
a safer bet, since we can never be sure
about the future correctness of a more specific ToE.
%
An a priori argument in favor of UToE is as follows: What is the
best candidate for a ToE before i.e.\ in absence of any
observations? If somebody (but how and who?) would tell us that the
universe is computable but nothing else, universal self-sampling
seems like a reasonable a priori UToE.

\paradot{Comparison to anthropic self-sampling}
Our universal self-sampling assumption is related to anthropic
self-sampling \cite{Bostrom:02} but crucially different. The
anthropic self-sampling assumption states that a priori you are
equally likely any of the (human) observers in our universe.
First, we sample from any universe and any location (living or dead)
in the multiverse and not only among human (or reasonably
intelligent) observers. Second, we have no problem of what counts as
a reasonable (human) observer. Third, our principle is completely
formal.

Nevertheless the principles are related since (see inclusion of $s$)
{\em given} $o_{1:t}^{true}$ we also sample from the set of
reasonable observers, since $o_{1:t}^{true}$ includes snapshots of
other (human) observers.

\paradot{No Free Lunch (NFL) myth}
Wolpert \cite{Wolpert:97} considers algorithms for finding the
minimum of a function, and compares their average performance. The
simplest performance measure is the number of function evaluations
needed to find the global minimum. The average is taken uniformly
over the set of all functions from and to some fixed finite domain.
Since sampling uniformly leads with (very) high probability to a
totally random function (white noise), it is clear that on average
no optimization algorithm can perform better than exhaustive search,
and no reasonable algorithm (that is one that probes every function
argument at most once) performs worse. That is, all reasonable
optimization algorithms are equally bad on average. This is the
essence of Wolpert's NFL theorem and all variations thereof I am
aware of, including the ones for less uniform distributions.

While NFL theorems are cute observations, they are obviously irrelevant,
since nobody cares about the maximum of white noise functions.
Despite NFL being the holy grail in some research communities, the
NFL myth has little to no practical implication \cite{Stork:01}.

An analogue of NFL for prediction would be as follows: Let
$u_{1:n}\in\{0,1\}^n$ be uniformly sampled, i.e.\ the probability of
$u_{1:n}$ is $\lambda(u_{1:n})=2^{-n}$. Given $u_{1:t}^{true}$ we
want to predict $u_{t+1:n}$.
Let $u_{t+1:n}^p$ be any deterministic prediction. It is clear that
all deterministic predictors $p$ are on average equally bad (w.r.t.\
symmetric performance measures) in predicting uniform noise
($\lambda(u_{t+1:n}^p|u_{1:t}^{true})=2^{-(n-t)}$) .

How does this compare to the positive result under universal
self-sampling? There we also used a uniform distribution, but over
effective models=theories=programs. A priori we assumed all programs
to be equally likely, but the resulting universe distribution is far
from uniform. Phrased differently, we piped uniform noise (via
$M$, see below) through a universal Turing machine. We assume a universal
distribution $M$, rather than a uniform distribution $\lambda$.

Just assuming that the world has {\em any} effective structure breaks
NFL down, and makes Ockham's razor work \cite{Hutter:02ulaos}.
The assumption that the
world has {\em some} structure is as safe as (or I think even weaker than)
the assumption that e.g.\ classical logic is good for reasoning
about the world (and the latter one has to assume to make science
meaningful).

\paradot{Some technical details$^*$}
Readers not familiar with Algorithmic Information Theory
might want to skip this paragraph.
$P(u)$ in \req{eqMKm} tends for $L\to\infty$ to Solomonoff's a
priori distribution $M(u)$. In the definition of $M$
\cite{Solomonoff:64} only programs of length $=L$, rather than $\leq
L$ are considered, but since $\lim_{L\to\infty}{1\over
L}\sum_{l=1}^L a_l=\lim_{L\to\infty} a_L$ if the latter exists, they
are equivalent. Modern definitions involve a $2^{-l(q)}$-weighted
sum of prefix programs, which is also equivalent \cite{Li:08}.
Finally, $M(u)$ is also equal to the probability that a universal
monotone Turing machine with uniform random noise on the input tape
outputs a string starting with $u$ \cite{Hutter:04uaibook}. Further,
$l\equiv\Length(q_{min})=K\!m(u)$ is the monotone complexity of
$u:=u_{1:t}^{true}$. It is a deep result in Algorithmic Information
Theory that $K\!m(u)\approx-\lb M(u)$. For most $u$ equality holds
within an additive constant, but for some $u$ only within
logarithmic accuracy \cite{Li:08}. Taking the ratio of $M(u)\approx
2^{-K\!m(u)}$ for $u=u_{1:t}^{true}u_{t+1:n}$ and $u=u_{1:t}^{true}$
yields \req{eqMKm}.

The argument/result is not only technical but also subtle:
Not only are there $2^{L-l}$ programs equivalent to $q_{min}$ but
there are also ``nearly'' $2^{L-l}$ programs that lead to totally
different predictions. Luckily they don't harm probabilistic
predictions based on $P$, and seldomly affect deterministic
predictions based on $q_{min}$ in practice but can do so in theory
\cite{Hutter:06unimdlx}. One can avoid this problem by augmenting
Ockham's razor with Epicurus principle of multiple explanations,
taking all theories consistent with the observations but weigh them
according to their length. See \cite{Li:08,Hutter:04uaibook} for
details.

\newpage
\section{Discussion}\label{secDisc}

\paradot{Summary}
I have demonstrated that a theory that perfectly describes our
universe or multiverse, rather than being a Theory of Everything
(ToE), might also be a theory of nothing. I have shown that a
predictively meaningful theory can be obtained if the theory is
augmented by the localization of the observer. This resulted in a
truly Complete Theory of Everything (CToE), which consists of a
conventional (objective) ToE plus a (subjective) observer process.
Ockham's razor quantified in terms of code-length minimization has
been invoked to select the ``best'' theory (UCToE).

\paradot{Assumptions}
The construction of the subjective complete theory of everything
rested on the following assumptions:
$(i)$ The observers' experience of the world consists of a single
temporal binary sequence $o_{1:t}^{true}$.
All other physical and epistemological concepts are derived.
$(ii)$ There exists an objective world independent of any particular
observer in it.
$(iii)$ The world is computable, i.e.\ there exists an algorithm (a
finite binary string) which when executed outputs the whole
space-time universe. This assumption implicitly assumes (i.e.\
implies) that temporally stable binary strings exist.
$(iv)$ The observer is a computable process within the objective
world.
$(v)$ The algorithms for universe and observer are chosen at random,
which I called universal self-sampling assumption.

\paradot{Implications}
As demonstrated, under these assumptions, the scientific quest for a
theory of everything can be formalized.
As a side result, this allows to separate objective knowledge $q$ from
subjective knowledge $s$.
One might even try to argue that if $q$ for the best $(q,s)$ pair is
non-trivial, this is evidence for the existence of an objective
reality.
Another side result is that there is no hard distinction between
a universe and a multiverse; the difference is qualitative and semantic.
Last but not least, another implication is the validity of Ockham's
razor.

\paradot{Conclusion}
Respectable researchers, including Nobel Laureates, have dismissed
and embraced each single model of the world mentioned in Section
\ref{secSToE}, at different times in history and concurrently.
(Excluding All-a-Carte ToEs which I haven't seen discussed before.)
As I have shown, Universal ToE is the sanity critical point.

The most popular (pseudo) justifications of which theories are
(in)sane have been references to the dogmatic Bible, Popper's
limited falsifiability principle, and wrong applications of Ockham's
razor. This paper contained a more serious treatment of world model
selection. I introduced and discussed the usefulness of a theory in
terms of predictive power based on model {\em and} observer
localization complexity.

\newpage
\addcontentsline{toc}{section}{\refname}
\bibliographystyle{alpha}

\begin{small}

\end{small}

\newpage\appendix
\section{List of Notation}\label{secLN}

\begin{tabbing}
  \hspace{3cm} \= \hspace{11cm} \= \kill
  G,H,E,P,S,C,M,U,R,A,... \> \hspace{1.5cm} specific models/theories defined in Section \ref{secSToE}         \\[0.5ex]
  $T$                \> $\in\{$G,H,E,P,S,C,M,U,R,A,...$\}$ theory/model                      \\[0.5ex]
  ToE                \> Theoy of Everything (in any sense)                                   \\[0.5ex]
  ToE candidate      \> a theory that might be a partial or perfect or wrong ToE             \\[0.5ex]
  UToE               \> Universal ToE                                                        \\[0.5ex]
  CToE               \> Complete ToE (i+e+l+n+o)                                             \\[0.5ex]
  UCToE              \> Universal Complete ToE                                               \\[0.5ex]
  theory             \> model which can explain$\approx$describe$\approx$predict$\approx$compress observations \\[0.5ex]
  universe           \> typically refers to visible/observed universe                        \\[0.5ex]
  multiverse         \> un- or only weakly connected collection of universes                 \\[0.5ex]
  predictive power   \> precision and coverage                                               \\[0.5ex]
  precision          \> the accuracy of a theory                                             \\[0.5ex]
  coverage           \> how many phenomena a theory can explain/predict                      \\[0.5ex]
  prediction         \> refers to unseen, usually future observations                        \\[0.5ex]
  computability assumption:  that our universe is computable                                 \\[0.5ex]
  $q^T\in\{0,1\}^*$  \> the program that generates the universe modeled by theory $T$        \\[0.5ex]
  $u^q\in\{0,1\}^\infty$ \> the universe generated by program $q$: $\quad u^q=\UTM(q)$       \\[0.5ex]
  $\UTM$             \> Universal Turing Machine                                             \\[0.5ex]
  $s\in\{0,1\}^*$    \> observation model/program. Extracts $o$ from $u$.                    \\[0.5ex]
  $o^{qs}\in\{0,1\}^\infty$ \> Subject's $s$ observations in universe $u^q$: $\quad o^{qs}=\UTM(s,u^q)$ \\[0.5ex]
  $o_{1:t}^{true}$   \> True past observations                                               \\[0.5ex]
  $\breve q,\breve u$ \> Program and universe of UToE                                        \\[0.5ex]
  $S(o|u)$           \> Probability of observing $o$ in universe $u$                         \\[0.5ex]
  $Q(u)$             \> Probability of universe $u$ (according to some prob.\ theory $T$)    \\[0.5ex]
  $P(o)$             \> Probability of observing $o$                                         \\[0.5ex]
  $ $                \>                                                                      \\[0.5ex]
  $ $                \>                                                                      \\[0.5ex]
\end{tabbing}

\end{document}